\documentclass[11pt,letterpaper]{article}

\usepackage[english]{babel}
\usepackage[utf8]{inputenc}
\usepackage{amsmath}
\usepackage{amsfonts}
\usepackage{graphicx}
\usepackage{float}
\usepackage[colorinlistoftodos]{todonotes}
\usepackage[margin=1in]{geometry}
\usepackage{wrapfig}
\usepackage[justification=centering]{caption}
\usepackage{hyperref}

\title{{\huge Network Models in Epidemiology} \\ \vspace{12pt} Considering Discrete and Continuous Dynamics}
\author{Edward Rusu \\ University of Washington \\ email \href{mailto:epr24@uw.edu}{epr24@uw.edu}}
\date{24 April 2014}

\begin{document}
\maketitle

\begin{abstract}
Discrete and Continuous Dynamics is the first in a series of articles on Network Models for Epidemiology. This project began in the Fall quarter of 2014 in my continuous modeling course. Since then, it has taken off and turned into a series of articles, which I hope to compile into a single report. The purpose of the report is to explore mathematical epidemiology. In this article, we discuss the historical approach to disease modeling with compartmental models. We discuss the issues and benefits of using network models. We build a discrete dynamical system to describe infection and recovery of individuals in the population. Lastly, we detail the computational scheme for iterating this model.
\end{abstract}

\section{Mathematical Epidemiology and the SIR Model}

\subsection{Mathematical Epidemiology}

Diseases are studied in many areas of research, such as virology and  pathology. Epidemiology concerns itself with the onset and spread of a disease through a population. It considers disease characteristics, such as contagiousness, reproduction, and lethality; as well as population characteristics, such as susceptibility and connectivity. Furthermore, epidemiologists work to contain and eradicate the disease. It is not enough to only know how a disease spreads. Epidemiology also asks the question, ``How can we best stop the disease?'' In this, we consider control protocols, quarantine, and vaccination.

Mathematical Epidemiology applies the language of mathematics to the study of epidemiology. We apply mathematical theory to describe the disease within the population. We analyze for qualitative understanding, and we build models that give quantitative results. One such model is the classical $SIR$ model, which provides us with the notion of the {\it basic reproductive number}. Besides this, mathematical modeling provides us equations that describe the dynamics of infection and provide insight for effective control protocols.

Mathematical modeling is an essential tool for framing the world around us. It provides us with helpful insights and accurate guidance. We will examine the $SIR$ model, and then provide a network model to examine the spatial constraints of social structures.

\begin{figure}
\includegraphics[width = \textwidth]{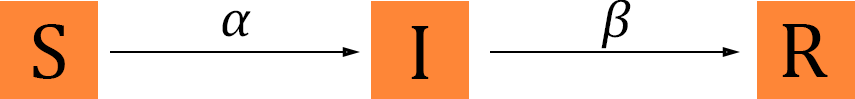}
\caption{Classes and Flows of the SIR model.}
\label{intro:SIR:model:fig}
\end{figure}

\subsection{SIR Model}

The $SIR$ model separates a population of individuals into three classes: Susceptible, Infectious, and Recovered. The model assumes that disease dynamics occur on a faster timescale than population dynamics, which results in a constant population $N$. Individuals flow from the Susceptible class into the Infectious class at a rate $\alpha$. For this flow, we employ the mass-action assumption, which assumes that susceptible individuals become infected via interactions with infectious individuals. Afterwards, individuals flow from the infectious class to the recovered class at a rate $\beta$; this dynamic occurs proportional to the amount of infectious individuals. That is, an individual's recovery does not depend on his surroundings but on his state of infection. In many cases, the recovered class contains individuals who have fully recovered from the disease (and thus become immune to it) and those who have died from the disease.

Figure \eqref{intro:SIR:model:fig} shows a flowchart for $SIR$, and Equation \eqref{intro:SIR:model:eq} gives a system of differential equations to model the flow. The model contains two parameters, namely, the flow rates.

\begin{equation} \label{intro:SIR:model:eq}
	\begin{aligned}
    \frac{dS}{dt} & = \frac{-\alpha SI}{N}\\
    \frac{dI}{dt} & = \frac{\alpha SI}{N} - \beta I\\
    \frac{dR}{dt} & = \beta I
    \end{aligned}
\end{equation}

\begin{figure}
\includegraphics[width = \textwidth]{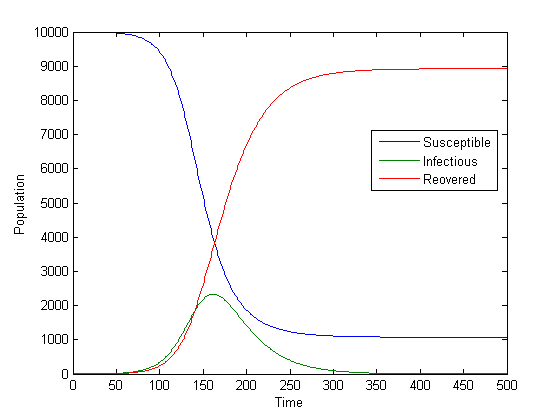}
\caption{Population levels in each class of the $SIR$ model with $\alpha = 0.1$ and $\beta = 0.04$. Initially, one person in a population of 10,000 is infected.}
\label{intro:SIR:xmpl:fig}
\end{figure}

Immediately, we see that Equation \eqref{intro:SIR:model:eq} is nonlinear. While we cannot solve each variable explicitly in time, there are ways to solve them with respect to one another. We do not go into those techniques here, but the readers should consult \cite{mathepid}. Instead, we apply numerical methods to solve this equation. In particular, we use Matlab's $ode45$ (Appendix A). Figure \eqref{intro:SIR:xmpl:fig} shows an example of the $SIR$ model with $\alpha = 0.1$ and $\beta = 0.04$. The total population is $N = 10,000$, and there is initially one infected individual within an entirely susceptible population.

As stated before, mathematical models provide us with quantitative results. One such result that has proved extremely useful is the basic reproductive number $R_0$. This number measures the epidemic potential of a disease. It is calculated under the following conditions: consider an entirely susceptible population with a single infected individual. This individual infects others at a rate of $\alpha$, and he will remain infected for $1/\beta$ time units. We say

\begin{equation*}
R_0 = \frac{\alpha}{\beta}
\end{equation*}

\noindent Essentially, $R_0$ tells us how many individuals an infected individual will infect. In the $SIR$ model, we say that an epidemic cannot occur unless $R_0 > 1$.

Consider the case shown in Figure \eqref{intro:SIR:xmpl:fig}. Here, $R_0 = 2.5$. At the start of the epidemic, every infected individual will infect two and half more individuals, so that the disease will spread exponentially and an epidemic occurs. On the other hand, consider $R_0 = 0.5$. Every two infected individuals will only infect a single individual, so the disease will decay exponentially, and an epidemic does not occur.

\begin{figure}
\begin{minipage}{0.5\textwidth}
\vspace*{12pt}
\centering
\begin{tabular}{c | c}
    Disease & $R_0$ \\
    \hline
    Measles & 12-18 \\
    Rubella & 5-7 \\
    Small Pox & 5-7 \\
    HIV & 2-5 \\
    Ebola & 1.5-2.5
    \end{tabular}
    \captionof{table}{Infectious diseases and their correlating basic reproductive numbers $R_0$ \cite{R0}.}
    \label{intro:SIR:R0:table}
\end{minipage}
\begin{minipage}{0.5\textwidth}
\centering
\includegraphics[width = 0.5\textwidth]{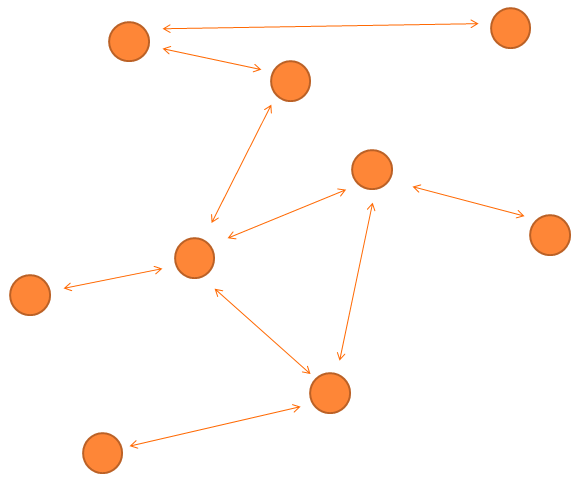}
\captionof{figure}{Network with individuals (nodes) and connections (edges).}
\label{intro:network:fig}
\end{minipage}
\end{figure}

Table \eqref{intro:SIR:R0:table} shows several infectious diseases and their correlating basic reproductive number $R_0$. An individual infected with Measles will infect between 12 and 18 other individuals before recovering from the infection! Measles, Rubella, and Small Pox are extremely infectious, and the CDC outlines vaccination requirements for each state \cite{vaccine}. Consider Ebola with an $R_0$ of 2. Even though its value is relatively low, it is currently causing in epidemic in West Africa. The Ebola epidemic is an example of why mathematical epidemiology is so important.

So we see that using mathematical modeling has provided us with useful information about the spread of disease. Equation \eqref{intro:SIR:model:eq} provides us with a basic understanding of the dynamics of infection. We have parameters $\alpha$ and $\beta$ to tune for data fitting and predictions. We even have the basic reproductive number to give us an epidemic threshold.

\subsection{Deterministic Compartmental Models}

There are many alterations and modifications to the $SIR$ model, such as the $SIS$ model, which assumes the infection does {\it not} confer immunity; the $MSIR$, which assumes that some are naturally immune to the infection (useful for modeling the spread of Measles); the $SEIR$, which also classifies some individuals as exposed but not yet infectious; and many more models. Each of these models is a deterministic compartmental model that separates the population into classes, and the flows between these classes are modeled with a low-dimensional system of ordinary differential equations.

While these models certainly have their merit, they exhibit a flawed assumption: the mass-action assumption. These models assume a well-mixed homogeneous population; that is, every individual in the population is in contact with every other individual. We intuitively know this to be false. Societies are not random and well-mixed but are built in social structures. For example, you are much more likely to be infected by a coworker, friend, or family member than you are by a random individual in the population because you spend much more time and have closer contact with coworkers, friends, and family members. Compartmental models do not account for these social structures. Furthermore, compartmental models assume that every individual in the population infects and recovers at the same rate. In reality, individuals differ in their contagiousness, susceptibility, recovery, immunity, etc. We hope to build a model that utilizes the basic notions of the $SIR$ model while considering social structures and individuality.

\section{Network Models}

For our purposes, we utilize a network to describe the spread of the infection. The nodes will be the individuals and the edges will be the probability that infection is transmitted. This probability accounts for characteristics such as susceptibility and contagiousness. Since each individual has unique characteristics, each edge is a different probability and is actually double-sided; that is, different value for different directions. Figure \eqref{intro:network:fig} shows such a network.

\subsection{Building and Analyzing a Network}

Network models are generally more difficult to analyze than compartmental models. Imagine building a network of students in a school. We want to determine how the network is connected. If we say that students are connected by classes, we could look up the schedules of every single student and build the network from this information. But students are also connected by extracurricular activities, so we would pull up the registry of clubs and sports to make further connections. Then we realize that students are most closely connected to their friends, so we hand out a questionnaire asking students to list each of their friends. Assuming the questionnaires are accurately completed, we take this data and build our network even further. While this is easily said, keep in mind that the average size of a high school in the United States in the year 2000 was 752 students \cite{School}! Imagine handing out and tracking down 752 questionnaires. Then imagine inputting all that data on a computer and creating the network. Imagine doing it for a system that is not as highly organized as a school and doesn't have registries of individuals. Before we get too lost in our imaginations, let us just say that collecting the information to build the network is a very involved process.

Suppose we have done all the required work, and we have our network. How do we analyze it? Let's begin by considering several metrics (or concepts) of the network:
\begin{itemize}
\item Connectedness: Is every individual connected to each other, or are there pockets of individuals isolated from the rest?
\item Distance: Individuals have many connections; what is the length of the shortest path between two individuals?
\item Diameter: What is the largest of these distances?
\item Connectivity: How many individuals are directly connected to a specific individual?
\item Degree Distribution: What is the distribution, if any, of the connectivity?
\end{itemize}

We've listed just a few concepts to explore within the network. We can also ask questions about population mixing, local clustering, etc. Note that there is a difference between the first three concepts and the last two. The first three are global concepts; we cannot know them unless we know information about the entire network. In contrast, the last two are local concepts, which we can know just by examining the individuals directly.

Ultimately, our goal is to understand how the network affects the spread of information (or in our case, infection). Once we have that understanding, we can determine how to slow the spread and even contain it. Already we see that the amount of effort to build the network and the analysis to understand it greatly outweighs what we did for the $SIR$ model. We motivate ourselves by remembering that compartmental models do not account for important factors that can be manipulated to contain and eradicate infections.

\subsection{Fundamental Networks}

\begin{figure}
    \begin{minipage}{0.24\textwidth}
	\includegraphics[width = 0.95\textwidth]{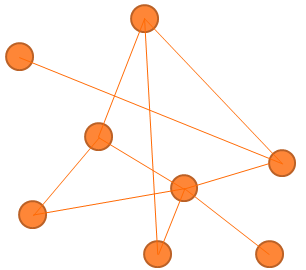}
    \captionof{figure}{Randomly connected network.}
    \label{intro:network:random:fig}
    \end{minipage}
    \begin{minipage}{0.24\textwidth}
	\includegraphics[width = 0.95\textwidth]{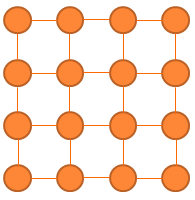}
    \captionof{figure}{Lattice network. Each individual connected to the four nearest individuals.}
    \label{intro:network:lattice:fig}
    \end{minipage}
    \begin{minipage}{0.24\textwidth}
	\includegraphics[width = 0.95\textwidth]{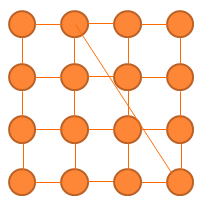}
    \captionof{figure}{Small world network. Mostly lattice with long-range connections throughout the network.}
    \label{intro:network:small_world:fig}
    \end{minipage}
    \begin{minipage}{0.24\textwidth}
	\includegraphics[width = 0.95\textwidth]{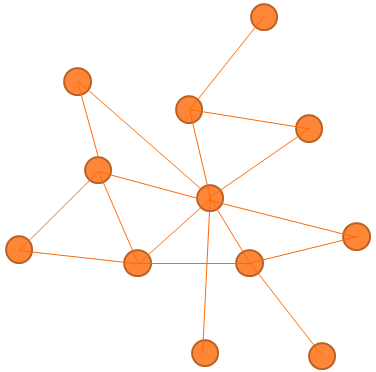}
    \captionof{figure}{Scale free network. New nodes tend to connect to already popular nodes.}
    \label{intro:network:scale_free:fig}
    \end{minipage}
\end{figure}

We have emphasized the need for in-depth data collection in order to thoroughly and accurately depict a population. For applications, this procedure is critical; inaccurate models due to poor data is often the reason why application problems fail, which leads to questioning the authority of the model \cite{leveque}. In this study, however, we want to be as general as possible. Remember that we are not yet considering an application; we are only exploring the effect spatial structure has on the spread of disease through a population network.

For our purposes, we will consider the four fundamental networks depicted in Figures (4-7). Figure \eqref{intro:network:random:fig} shows a random network. The nodes are randomly connected with no structure or pattern. As stated before, we know that individuals interact within social structures, so we expect there to be very limited randomness. We will examine it for qualitative understanding and not for practical application.

Figure \eqref{intro:network:lattice:fig} shows a lattice network. Each individual is connected to the closest individuals. Body individuals have four connections, edge individuals have three, and corner individuals have two. While the lattice structure may be practical for tightly organized scenarios (such as a complex of cubicles or buildings in a city block), it may be too strict to account for human movement and interactivity.

Figure \eqref{intro:network:small_world:fig} shows a small world network. This network is mostly lattice with several long-range connections that reduce the diameter the of the network. In a lattice network, it takes six connections to get from the top left to the bottom right. With small world, it only takes two.

Figure \eqref{intro:network:scale_free:fig} shows a scale free network. This network is also recognized by the saying ``rich get richer''. Nodes with high connectivity will attract new nodes and achieve even higher connectivity. Although it is the most complicated of the four, it most accurately describes human interactions.

Lastly, we note that the network itself is dynamic. That is, connections between individuals change over time. All of the data collection and analysis we perform to determine network characteristics must constantly be redone to keep the network up to date. In most cases, we would attempt to model the changing connections via a system of differential equations. Thus, we would have dynamical system that determines the connections in the network that determines the spread of infection through a population. In our model, we will assume that disease dynamics occur on a timescale much faster than network dynamics, so we will keep the network constant.

\section{Discrete Dynamical System}

\begin{figure}
\begin{minipage}{0.5\textwidth}
\includegraphics[width = 0.96\textwidth]{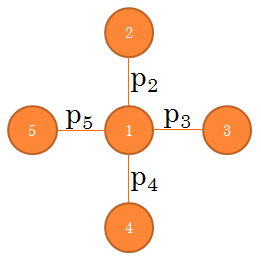}
\caption{Lattice network of five nodes.}
\label{dds:setup:fig}
\end{minipage}
\begin{minipage}{0.5\textwidth}
\vspace{17pt}
\includegraphics[width = 0.96\textwidth]{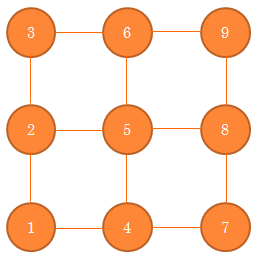}
\caption{3-by-3 lattice network.}
\label{dds:3x3:fig}
\end{minipage}
\end{figure}

Let us now consider how an infection spreads through a network population. We assume the infection is transmittable only via direct contact between individuals (not airborne and not environmental). Individuals experience two dynamics: infection and recovery.

To understand how the infection spreads, we will consider a lattice network of five nodes, as shown in Figure \eqref{dds:setup:fig}. Let $u_i^m$ be the probability that individual $i$ is infected at time $m$. Interactions happen in a discrete setting; individuals do not continuously interact with one another but do so as discrete events. Thus, our time variable $m$ measures discrete interactions, so $m \in \mathbb{N}$, where $\mathbb{N} = \{ 0, 1, 2, 3, ... \}$. For simplicity, we scale $m$ so that $\Delta m = 1$.

Consider individual 1 shown in Figure \eqref{dds:setup:fig}. Without any interactions, the probability of infection at the next time step should be what it was in the previous time step, so that

\begin{equation*}
u_1^{m+1} = u_1^m
\end{equation*}

Now consider that this individual interacts with potentially infectious individuals. We have probabilities $p_i$ unique to each interaction. For simplicity, we will assume that these probabilities are given from a normal distribution between 0 and 1. Furthermore, we will consider a parameter $s$ that scales these probabilities, changing an infection's contagiousness. For individual 1, we have

\begin{equation*}
u^{m+1}_1 = u^m_1 + s(p_2u^m_2 + p_3u_3^m + p_4u^m_4 + p_5u^m_5)
\end{equation*}

\noindent and in general we can write

\begin{equation} \label{dds:inf:model:eq}
u^{m+1}_i = u^m_i + s(p_{i+1}u^m_{i+1} + p_{i+2}u^m_{i+2} + p_{i+3}u^m_{i+3} + p_{i+4}u^m_{i+4})
\end{equation}

\begin{figure}
\centering
\includegraphics[width = 0.75\textwidth]{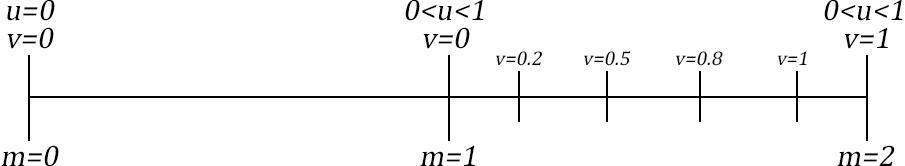}
\caption{Difference between continuous recovery and discrete interactions.}
\label{dds:cont-disc:fig}
\end{figure}

Let us now consider how an individual recovers. Let $v_i(t)$ be the probability of recovery for individual $i$ as a function of time $t$. Unlike infection, recovery dynamics occur continuously. However, we don't care about the recovery values for most of $t$. We actually only care about $v$ when $t = m$; that is, by the next interaction. Imagine an individual who goes to work and interacts with infectious individuals at $m = 1$. By the end of this interaction, this individual is himself infectious to within some probability, and his body is already at work fighting the infection. He goes home for the evening, watches a few episodes of the Walking Dead, sleeps eight hours, and wakes up the next morning for work. All the while, his body has been working to recover from the disease. But we don't care about what his recovery is between episodes, at 2am, or even when he sips his morning coffee. We only care if he recovers before his next interaction at $m = 2$ because that determines how the infection spreads. Thus, we can consider recovery values at discrete times $m$. An graphical example is shown in Figure \eqref{dds:cont-disc:fig}.

Suppose individual $i$ has a unique recovery constant $r_i$ that accounts for his body's ability to fight off the disease, environmental contributions to his health, medical assistance that effects his recovery, etc. We understand that recovery is a response to infection (the individual doesn't need to recovery if he is not infected). Thus,

\begin{equation} \label{dds:rec:model:eq}
v^{m+1}_i = v^m_i + r_iu^m_i
\end{equation}

As an example, consider this system over a 3-by-3 lattice, as shown in Figure \eqref{dds:3x3:fig}. For simplicity, we write the dynamics in matrix vector form. Let $p_{i,j}$ be the probability that individual $j$ infects individual $i$. For infection, we have

\begin{align*}
P \hspace{125pt} \vec{u}^m \hspace{6pt} & = \hspace{9pt} \vec{u}^{m+1} \\
	\begin{bmatrix}
    1 & p_{1,2} & 0 & p_{1,4} & 0 & 0 & 0 & 0 & 0 \\
    p_{2,1} & 1 & p_{2,3} & 0 & p_{2,5} & 0 & 0 & 0 & 0 \\
    0 & p_{3,2} & 1 & 0 & 0 & p_{3,6} & 0 & 0 & 0 \\
    p_{4,1} & 0 & 0 & 1 & p_{4,5} & 0 & p_{4,7} & 0 & 0 \\
    0 & p_{5,2} & 0 & p_{5,4} & 1 & p_{5,6} & 0 & p_{5,8} & 0 \\
    0 & 0 & p_{6,3} & 0 & p_{6,5} & 1 & 0 & 0 & p_{6,9} \\
    0 & 0 & 0 & p_{7,4} & 0 & 0 & 1 & p_{7,8} & 0 \\
    0 & 0 & 0 & 0 & p_{8,5} & 0 & p_{8,7} & 1 & p_{8,9} \\
    0 & 0 & 0 & 0 & 0 & p_{9,6} & 0 & p_{9,8} & 1
    \end{bmatrix}
    \begin{bmatrix}
    u^{m}_1 \\ u^{m}_2 \\ u^{m}_3 \\ u^{m}_4 \\ u^{m}_5 \\ u^{m}_6 \\ u^{m}_7 \\ u^{m}_8 \\ u^{m}_9
    \end{bmatrix}
    & = 
    \begin{bmatrix}
    u^{m+1}_1 \\ u^{m+1}_2 \\ u^{m+1}_3 \\ u^{m+1}_4 \\ u^{m+1}_5 \\ u^{m+1}_6 \\ u^{m+1}_7 \\ u^{m+1}_8 \\ u^{m+1}_9
    \end{bmatrix}
\end{align*}

\noindent $s$ would normally multiply the interaction terms (the off-diagonals), but here we have a assumed $s = 1$. The main diagonal of $P$ holds the $u^m_i$ values, and the off-diagonals record the interactions. If we change the network, the main diagonal of $P$ remains the same, but the off-diagonals will change. The rows tell how a certain individual becomes infected, and the columns tell how that individual infects others.

Recovery is even easier to vectorize because it simply involves an element-wise multiplication between $\vec{r}$ and $\vec{u}^m$.

\begin{align*}
\vec{v}^m \hspace{8pt} + \hspace{8pt} \vec{r} \hspace{27pt} \vec{u}^m \hspace{6pt} & = \hspace{9pt} \vec{v}^{m+1} \\
    \begin{bmatrix}
    v^{m}_1 \\ v^{m}_2 \\ v^{m}_3 \\ v^{m}_4 \\ v^{m}_5 \\ v^{m}_6 \\ v^{m}_7 \\ v^{m}_8 \\ v^{m}_9
    \end{bmatrix}
    + 
    \begin{bmatrix}
    r_1 \\ r_2 \\ r_3 \\ r_4 \\ r_5 \\ r_6 \\ r_7 \\ r_8 \\ r_9
    \end{bmatrix}
    \cdot
    \begin{bmatrix}
    u^{m}_1 \\ u^{m}_2 \\ u^{m}_3 \\ u^{m}_4 \\ u^{m}_5 \\ u^{m}_6 \\ u^{m}_7 \\ u^{m}_8 \\ u^{m}_9
    \end{bmatrix}
    & = 
    \begin{bmatrix}
    v^{m+1}_1 \\ v^{m+1}_2 \\ v^{m+1}_3 \\ v^{m+1}_4 \\ v^{m+1}_5 \\ v^{m+1}_6 \\ v^{m+1}_7 \\ v^{m+1}_8 \\ v^{m+1}_9
    \end{bmatrix}
\end{align*}

\subsection{Computational Iteration}

All of the insight we gather from the network model will be a result of numerical simulation. Thus, rather than simply presenting the Matlab code for iterating the dynamics in Equations \eqref{dds:inf:model:eq} and \eqref{dds:rec:model:eq}, we will discuss how to handle certain caveats. Here, we focus on the iteration scheme. Suppose the infection matrix $P$ has already been created for some particular network. We start by setting the time stepping loop.

\begin{verbatim}
% Discrete Dynamical System Iteration
for i = 1:sims
    v(:,i+1) = v(:,i) + r0.*u(:,i); % Iterate Recovery
    u(:,i+1) = u(:,i) + P*u(:,i); % Iterate Infection
end
\end{verbatim}

\noindent where $sims$ is the number of time steps to simulate.

While this is a good start, it is not enough. The values for $u$ and $v$ will very quickly grow larger than 1. This will inaccurately cause the infection to become more contagious and recovery to speed up. We correct these values in the loop.

\begin{verbatim}
for i = 1:sims
    % Correct probabilities
    nduGE1 = find(u(:,i) >= 1);
    ndvGE1 = find(v(:,i) >= 1);    
    u(nduGE1,i) = 1;
    v(ndvGE1,i) = 1;
    
    % Iterate 
    v(:,i+1) = v(:,i) + r0.*u(:,i); % Iterate Recovery
    u(:,i+1) = u(:,i) + P*u(:,i); % Iterate Infection
end
\end{verbatim}

Our dynamics are starting to look better, but we haven't really accounted for recovery in all that it means. When an individual recovers, he can no longer become infected, nor can he infect others. Initially, we might think to change the matrix $P$, setting the rows and columns for recovered individual to zero. However, modifying the $P$ matrix in each iteration greatly slows down our algorithm.

Instead, we include a control vector $\vec{h}$. While individual $i$ is susceptible, $h_i = 1$, and if individual $i$ is recovered, then $h_i = 0$. In each iteration, we element-wise multiply $\vec{h}$ with $\vec{u}$ before multiplying with $P$. This makes it so that the individual cannot infect others. After multiplying with $P$, we element-wise multiply the resulting vector again by $\vec{h}$. This makes it so that the individual can no longer become infected. (The last multiplication isn't actually necessary. We include it so that we can keep the infection numbers consistent).

Now we have a control parameter $\vec{h}$ that turns the infection on and off for certain individuals. The last thing to clarify is how $h_i$ changes from 0 to 1. Initially, we might guess that $h_i$ should become 0 at whatever time step $v^m_i = 1$. While this may seem intuitive, we must remember that we have two fundamentally different kinds of dynamics for infection and recovery: once is discrete, the other is continuous. Look again at Figure \eqref{dds:cont-disc:fig}: at $m = 1$, $v = 0$, so we would leave $h_i = 1$. However, the individual recovers before his next interaction, so that at $m = 2$, he should not have infection dynamics. Thus, we don't change $\vec{h}$ according to $\vec{v}^m$ but according to $\vec{v}^{m+1}$.

At this point, we have our final result for how we will iterate our dynamics.

\begin{verbatim}
for i = 1:sims
    % Correct Probabilities
    nduGE1 = find(u(:,i) >= 1);
    ndvGE1 = find(v(:,i) >= 1);    
    u(nduGE1,i) = 1;
    v(ndvGE1,i) = 1;    
    
    % Iterate v
    v(:,i+1) = v(:,i) + h.*(r0.*u(:,i));
    ndvGE1_next = find(v(:,i+1) >= 1);
    h(ndvGE1_next) = 0;
    
    % Iterate u
    u(:,i+1) = u(:,i) + h.*(P*(h.*u(:,i)));
end
\end{verbatim}

\section{Conclusion}

We have discussed several important characteristics of Deterministic Compartmental Models. We pointed out the flaw in the mass-action assumption, which led us to using network models to handle social structures and individual differences. While network models are significantly harder to create and analyze, they offer a much higher resolution for epidemiology dynamics.

We spent the last half of this article discussing the differences between discrete- and continuous-time dynamics found in infection and recovery. With some careful consideration, we were able to reduce recovery dynamics to a discrete setting and develop a fast and accurate scheme for iterating the system. Future articles will apply this integrator to the fundamental network models discussed above. In each case, we will draw mathematical analysis, discuss assumptions, and design solutions for infection containment.

\newpage

\section{Appendix A} \label{AppA}

\subsection{SIR Integrator} \label{AppA:SIR}

\begin{verbatim}
[t,y] = ode45(@(t,y) SIR(t,y,0.1,0.04,10000),1:500,[9999;1;0]);

figure(1)
plot(t,y)
xlabel('Time'), ylabel('Population')
legend('Susceptible','Infectious','Reovered','Location','Best')
\end{verbatim}

\begin{verbatim}
function dy = SIR(t,y,a,b,N)

dy = [-a*y(1)*y(2)/N;
    a*y(1)*y(2)/N - b*y(2);
    b*y(2)];
end
\end{verbatim}


\begin{thebibliography}{9}
\bibitem{mathepid}
Brauer, Fred, et al. \textit{Mathematical Epidemiology}. Berlin: Springer, 2008.
\bibitem{vaccine}
"State School and Childcare Vaccination Laws." Centers for Disease Control and Prevention. Centers for Disease Control and Prevention, 23 Mar. 2015. Web. 24 Apr. 2015.
\bibitem{R0}
History and Epidemiology of Global Smallpox Eradication. "Smallpox: Disease, Prevention, and Intervention". The CDC and the World Health Organization. Slide 16-17.
\bibitem{School}
"Overview of Public Elementary and Secondary Schools and Districts." National Center for Education Statistics. U.S. Department of Education, Sept. 2001. Web. Apr. 2015.
\bibitem{leveque}
LeVeque, Randall. "Tsunami Modeling." Princeton Companion to Applied Mathematics (n.d.): n. pag. Print.
\end{thebibliography}
\end{document}